\input{aipcheck}

\documentclass[
    ,final            
  ]
  {aipproc}

\layoutstyle{6x9}

\begin{document}

\title{Coll Positioning systems:\\ a two-dimensional approach}



\author{Joan Josep Ferrando}{
  address={Departament
d'Astronomia i Astrof\'{\i}sica, Universitat de Val\`encia, 46100
Burjassot, Val\`encia, Spain} }


\begin{abstract}
The basic elements of Coll positioning systems (n clocks
broadcasting electromagnetic signals in a n-dimensional space-time)
are presented in the two-dimensional case. This simplified approach
allows us to explain and to analyze the properties and interest of
these relativistic positioning systems. The positioning system
defined in flat metric by two geodesic clocks is analyzed. The
interest of the Coll systems in gravimetry is pointed out.
\end{abstract}

\maketitle

\section{INTRODUCTION}

The relativistic positioning systems were introduced by B. Coll a
few years ago at the Spanish Relativity Meeting celebrated in
Valladolid \cite{coll-1}. Remember that these systems are defined by
four clocks broadcasting their proper time. Here we will name them
for short 'Coll systems'. In a 'long contribution' to these
proceedings B. Coll explains the interest, characteristics and good
qualities of these relativistic positioning systems in the generic
four-dimensional case. In this short communication we present a
two-dimensional approach to the Coll systems.

The two-dimensional approach should help us understand better how
these relativistic systems work and the richness of the elements of
the Coll systems. Indeed, the simplicity of the 2-dimensional case
allows us to use precise and explicit diagrams which improve the
qualitative comprehension of the positioning systems. Moreover,
two-dimensional examples admit simple and explicit analytic results.

Nevertheless, it is worth remarking that the two-dimensional case
has particularities and results that cannot be generalized to the
generic four-dimensional case. Consequently, the two-dimensional
approach is suitable for learning basic concepts about positioning
systems, but it does not allow us to study some specific positioning
features that necessarily need a three- or a four-dimensional
approach.

In the second section we introduce the basic elements of a Coll
system: emission coordinates and its other essential physical
components. In the third section we explain the analytic method to
obtain the emission coordinates from an arbitrary null coordinate
system and we use it to develop the positioning system defined in
flat space-time by two geodesic clocks. We finish with some comments
about other positioning subjects that we are considering at present.

\section{BASIC CONCEPTS AND ESSENTIAL PHYSICAL COMPONENTS}

In a two-dimensional space-time, let $\gamma_1$ and $\gamma_2$ be
the world lines of two clocks measuring their proper times $\tau^1$
and $\tau^2$ respectively. Suppose they broadcast them by means of
electromagnetic signals, and that these signals reach each other of
the world lines. The future light cones (here reduced to pairs of
'light' lines) cut in the region between both emitters and they are
tangent outside. Thus, these proper times do not distinguish
different events on the emission null geodesics of the exterior
region.

The internal region, bounded by the emitter world lines, defines a
coordinate domain, the {\em emission coordinate domain} $\Omega$.
Indeed, every event on this domain can be distinguished by the times
$(\tau^1,\tau^2)$ received from the emitter clocks. In other words,
the past light cone of every event on the emission domain cuts the
emitter world lines at $\gamma_1(\tau^1)$ and $\gamma_2(\tau^2)$
respectively: then $\{\tau^1,\tau^2\}$ are the {\it emission
coordinates} of this event. An important property of the emission
coordinates we have defined is that they are null coordinates. The
plane $\{\tau^1\}\times\{\tau^2\}$ in which the different data of
the positioning system can be transcribed is called the {\em grid}
of the positioning system.

An observer $\gamma$ travelling throughout the emission coordinate
domain and equipped with a receiver which allows to read the proper
times $(\tau^1,\tau^2)$ at each point of his trajectory is a user of
this positioning system.

In defining the emission coordinates we have introduced the first
essential physical components of a Coll system:

\begin{itemize}
\item [-]
The principal emitters $\gamma_1$, $\gamma_2$, which broadcast their
proper time $\tau^1$, $\tau^2$.
\item [-]
The users $\gamma$, travelling in the emitter coordinate domain
$\Omega$, receive the emitted times $\{\tau^1, \tau^2\}$ (their
emitter coordinates).
\end{itemize}

These elements define a generic, free and immediate location system
(it can be defined in a generic space-time; it can be defined
without knowing the gravitational field; a user knows his
coordinates without delay).

Any user receiving continuously the {\em user's positioning data}
$\{\tau^1, \tau^2\}$ may extract his trajectory, $\tau^2 =
F(\tau^1)$, in the grid. Nevertheless, whatever the user be, these
data are insufficient to construct both of the two emitter
trajectories.

In order to give to any user the capability of knowing the emitter
trajectories in the grid, the positioning system must be endowed
with a device allowing every emitter to also broadcast the proper
time it is receiving from the other emitter:
\begin{itemize}
\item[-] The emitters $\gamma_1$, $\gamma_2$ are also
transmitters: they receive the signals (such as a user) and
broadcast them.
\item[-] The users $\gamma$ also receive the transmitted times $\{\bar{\tau}^1,
\bar{\tau}^2\}$.
\end{itemize}

In other words, the clocks must be allowed to broadcast {\em their
emission coordinates} and then, any user receiving continuously the
{\it emitter's positioning data}  $\{\tau^1, \tau^2; \bar{\tau}^1,
\bar{\tau}^2\}$ may extract from them the equations $\bar{\tau}^2 =
\varphi_1(\tau^1)$ and $\bar{\tau}^1 = \varphi_2(\tau^2)$ of the
emitter trajectories. A positioning system so endowed will be called
an {\em auto-located positioning system}.

Eventually, the positioning system can be endowed with complementary
devices. For example, in obtaining the dynamic properties of the
system:
\begin{itemize}
\item[-] The emitters $\gamma_1$, $\gamma_2$ can
carry accelerometers and broadcast their acceleration.
\item[-] The users $\gamma$ can also receive the emitter acceleration data $\{\alpha_1,
\alpha_2\}$.
\end{itemize}

In some cases, it can be useful that the users generate their own
data: they can carry a clock that measures their proper time $\tau$
and an accelerometer that measures their acceleration $\alpha$.

Thus, a Coll positioning system can be performed in such a way that
any user can obtain a subset of the user data: $ \{\tau^1, \tau^2;
\bar{\tau}^1, \bar{\tau}^2; \alpha_1, \alpha_2; \tau, \alpha\} $.

\section{POSITIONING WITH GEODESIC EMITTERS IN FLAT METRIC}

Let us assume the {\it proper time history of two emitters} to be
known in a null coordinate system $\{\texttt{u},\texttt{v}\}$:
\begin{equation}  \label{principalemit}
\gamma_1 \equiv \cases{\texttt{u} = u_1(\tau^1) \cr \texttt{v} =
v_1(\tau^1)} \qquad \qquad \gamma_2 \equiv \cases{\texttt{u} =
u_2(\tau^2) \cr \texttt{v} = v_2(\tau^2)}
\end{equation}
We can introduce the proper times as coordinates $\{\tau^1,\tau^2\}$
as follows:

\begin{equation}  \label{coordinatechange0}
\texttt{u} = u_1(\tau^1) \, , \qquad \qquad  \texttt{v} =
v_2(\tau^2)
\end{equation}
This change defines {\it emission null coordinates} in the {\it
emission coordinate domain} $\Omega \equiv  \big\{ (\texttt{u},
\texttt{v}) \ /  \quad F_2^{-1}(\texttt{v}) \leq \texttt{u} \, ,
\quad  F_1(\texttt{u}) \leq \texttt{v} \big\} $. In the region
outside $\Omega$ this change also determines null coordinates which
are an extension of the emission coordinates. But in this region the
coordinates are not physical, i.e. are not the emitted proper times
of the {\it principal emitters} $\gamma_1$, $\gamma_2$.

Now we use this procedure for the case of two {\it geodesic}
emitters $\gamma_1$, $\gamma_2$ in {\it flat space-time}. In
inertial null coordinates $\{\textrm{u},\textrm{v}\}$ the proper
time parametrization of the emitters are:
\begin{equation}  \label{inertialprincipalemit}
\gamma_1 \equiv \cases{ \, \texttt{u} = \lambda_1 \tau^1 \cr \,
\texttt{v} = \frac{1}{\lambda_1} \tau^1 + v_0 } \qquad \quad
\gamma_2 \equiv \cases{ \, \texttt{u} =  \lambda_2 \tau^2 + u_0 \cr
\, \texttt{v} = \frac{1}{\lambda_2} \tau^2 }
\end{equation}
Then, the emitter coordinates $\{\tau^1,\tau^2\}$ are defined by the
change:
\begin{equation}  \label{coordinatechange0}
\texttt{u}  = u_1(\tau^1) = \lambda_1 \tau^1 \, , \quad \qquad
\texttt{v} = v_2(\tau^2) = \frac{1}{\lambda_2} \tau^2
\end{equation}

From here we can obtain the metric tensor in emitter coordinates
$\{\tau^1,\tau^2\}$ and we obtain: $\, ds^2 = \lambda \, d \tau^1 d
\tau^2 \, , \,  \lambda \equiv \frac{\lambda_1}{\lambda_2}$. On the
other hand, in emission coordinates $\{\tau^1,\tau^2\}$, the
equations of the emitter trajectories are:
\begin{equation}  \label{emit-taus-inertial-1}
\gamma_1 \equiv \cases{ \tau^1 = \tau^1 \cr \tau^2 =
\varphi_1(\tau^1) \equiv  \frac{1}{\lambda} \tau^1 + \tau^2_0 }
\qquad \gamma_2 \equiv \cases{ \tau^1 = \varphi_2(\tau^2) \equiv
\frac{1}{\lambda} \tau^2 + \tau^1_0 \cr \tau^2 = \tau^2 }
\end{equation}

Let $\gamma$ be a user of this positioning system. What information
can this user obtain from the public data? Evidently $(\tau^1,
\tau^2)$ place the user on the user grid, and $(\bar{\tau}^1,
\bar{\tau}^2)$, $\bar{\tau}^i = \varphi_j(\tau^j)$, place the
emitters on the user grid. On the other hand, the metric component
could be obtained from the emitter's positioning data $\{\tau^1,
\tau^2; \bar{\tau}^1, \bar{\tau}^2\}$ at two events. The space-time
interval is:
\begin{equation}
ds^2 = \sqrt{\frac{\Delta\tau^1 \Delta\tau^2}{\Delta\bar{\tau}^1
\Delta\bar{\tau}^2}} \, d \tau^1 d \tau^2
\end{equation}

\section{DISCUSSION AND WORK IN PROGRESS}

We finish this talk with some comments about other positioning
subjects we are studying at present. Firstly, the interest of the
Coll systems in gravimetry. If we suppose that the user has no
previous information on the gravitational field, what metric
information can a user obtain from the public and proper user data?
Can a user do gravimetry by using our positioning system? We have
shown that \cite{cfm-1}:

- The public data $\{\tau^1, \tau^2; \bar{\tau}^1, \bar{\tau}^2;
\alpha_1, \alpha_2\}$ determine the space-time metric interval and
its gradient along the emitter trajectories.

- The public-user data $\{\tau^1, \tau^2; \tau, \alpha\}$ determine
the space-time metric interval and its gradient along the user
trajectory.

The development of a general method that offer a good estimation of
the gravitational field from this information is still an open
problem, but some preliminary results show its interest in
determining the parameters in a given (parameterized) model
\cite{cfm-2}.

On the other hand, some circumstances can lead to take another point
of view: the user knows the space-time in which he is immersed
(Minkowski, Schwarzschild,...) and we want to study the information
that the data received by the user offer. We have undertaken this
problem for the flat case an we have obtained interesting
preliminary results. In particular, we have shown that \cite{fm}:

- {\it If a user receives the emitter positioning data $\{\tau^1,
\tau^2; \bar{\tau}^1, \bar{\tau}^2\}$ along his trajectory and the
acceleration of one of the emitters during a sole {\em echo
interval} (i.e., travel time of a two-way signal from an emitter to
the other), then this user knows: his local unities of time and
distance, the metric interval in emission coordinates everywhere,
his own acceleration and the acceleration of the principal emitters,
the change between emission and inertial coordinates, his trajectory
and the emitter trajectories in inertial coordinates}.

\begin{theacknowledgments}
This work has been supported by the Spanish Ministerio de
Educaci\'on y Ciencia, MEC-FEDER project AYA2003-08739-C02-02.
\end{theacknowledgments}

\bibliographystyle{aipproc}   

\bibliography{sample}

\IfFileExists{\jobname.bbl}{}
 {\typeout{}
  \typeout{******************************************}
  \typeout{** Please run "bibtex \jobname" to optain}
  \typeout{** the bibliography and then re-run LaTeX}
  \typeout{** twice to fix the references!}
  \typeout{******************************************}
  \typeout{}
 }

\end{document}